\documentclass[manuscript]{acmart}
\usepackage{lipsum}
\AtBeginDocument{%
  \providecommand\BibTeX{{%
    \normalfont B\kern-0.5em{\scshape i\kern-0.25em b}\kern-0.8em\TeX}}}

\setcopyright{rightsretained}
\copyrightyear{2021}
\acmYear{2021}
\acmDOI{}

\acmConference[CHI '21]{Position paper at the workshop ``What Can CHI do About Dark Patterns"}{May 08, 2021}{Yokohama, Japan}
\acmBooktitle{Workshop: ``What Can CHI do About Dark Patterns"  at the CHI Conference on Human Factors in Computing Systems (CHI ’21), May 8–13, 2021, Yokohama Japan}
\acmPrice{15.00}
\acmISBN{978-1-4503-XXXX-X/18/06}

\begin{document}

\title{Dark Patterns, Electronic Medical Records, and the Opioid Epidemic}

\author{Daniel Capurro}
\authornote{Both authors contributed equally to this research.}
\email{dcapurro@unimelb.edu.au}
\orcid{0000-0002-9256-1256}
\author{Eduardo Velloso}
\email{eduardo.velloso@unimelb.edu.au}
\orcid{0000-0003-4414-2249}
\affiliation{%
  \institution{University of Melbourne}
  \city{Melbourne}
  \state{Victoria}
  \country{Australia}
  \postcode{3010}
}

\renewcommand{\shortauthors}{Capurro and Velloso}

\begin{abstract}
Dark patterns have emerged as a set of methods to exploit cognitive biases to trick users to make decisions that are more aligned with a third party than to their own. These patterns can have consequences that might range from inconvenience to global disasters. We present a case of a drug company and an electronic medical record vendor who colluded to modify the medical record's interface to induce clinicians to increase the prescription of extended-release opioids, a class of drugs that has a high potential for addiction and has caused almost half a million additional deaths in the past two decades. Through this case, we present the use and effects of dark patterns in healthcare, discuss the current challenges, and offer some recommendations on how to address this pressing issue.
\end{abstract}

\ccsdesc[500]{Human-centered computing~User interface design}
\ccsdesc[300]{Information systems~Decision support systems}
\ccsdesc[300]{Applied computing~Health informatics}
\ccsdesc[300]{Social and professional topics~Medical records}

\keywords{clinical decision support systems, electronic medical records, dark patterns, cognitive bias}


\maketitle

\section{Introduction}



The amount of information required to make sound clinical decisions is enormous and continuously growing~\cite{charlin2012clinical, TheScientist2002}. The combination of patient attributes, laboratory results, imaging---along with patient values and preferences---makes this process very complex~\cite{szulewski2017measuring}. Further, the availability of novel genetic and molecular assays that test for hundreds or thousands of genes or proteins and the emergence of previously unknown diseases make the task impossible without the support of external systems to aid clinicians and patients in sound decision making. The complexity of such decisions is one of the reasons explaining why patients only receive around half of the recommended health interventions \cite{runciman2012caretrack, mcglynn2003quality}---a situation with disastrous consequences for their health and well-being.

Electronic Medical Records (EMRs) have emerged in the past twenty years as comprehensive information systems used to collect and synthesize patient data, and to provide decision support for health professionals. The category of devices and artifacts used to facilitate clinical decision making are collectively known as \textit{clinical decision support systems} (CDSSs). CDSSs can facilitate the documentation of relevant clinical information, alert clinicians about abnormal laboratory results, suggest relevant clinical pathways, summarize patient variables, and many other forms of decision support. Although CDSSs can be implemented through non-digital methods, such as paper reminders \cite{pantoja2019manually}, most CDSSs are embedded in Electronic Medical Records. Given the diversity of clinical problems, interventions, and possible outcomes, evidence supporting the use of CDSSs is heterogeneous, but there is a growing number of patient and process outcomes that have been shown to be improved through the use of CDSSs. As an example, a recent overview of systematic reviews on the use of CDSS to improve outcomes in patients with diabetes found that 83\% of all included studies showed positive impacts on processes of care and 1/3 of them demonstrated benefits in managing blood pressure, blood glucose, and even a reduction in mortality \cite{jia2019evaluation}. The accumulating evidence has made CDSSs an attractive method to influence clinical decision making and to change clinician's behaviour.

However, at the same time that the digitisation of CDSSs has enhanced the speed, accuracy, and scalability of clinical decision making, it has also increased the risk of making the decision process more opaque and of reducing the agency of clinicians. This risk is amplified by recent advances in artificial intelligence and machine learning, which despite offering promising improvements in decision making performance, might not allow for inspection of how the recommendations were reached. This context, combined with competing interests from pharmaceutical companies and medical device manufacturers, creates fertile grounds for the proliferation of \textit{dark interface design patterns} in CDSSs. We consider dark patterns to be \textbf{common interface design solutions leveraging cognitive biases and heuristics to trick users into making decisions that are more aligned to third party interests than to their own}. In this paper we discuss a case of dark patterns influencing patient treatment through the modification of a CDSS embedded in a commercial electronic health record.

\section{Clinical Dark Pattern}

Chronic pain is a frequent and difficult to treat condition that can generate significant consequences to patients and to the healthcare system overall. Significant efforts have been made to train clinicians to identify and treat patients with chronic pain. A range of possible treatments for chronic pain exist, ranging from surgical implants to psychological therapies. Among them, the use of opioids has been long recommended for the treatment of severe to moderate pain related to cancer or other serious illnesses~\cite{world1986cancer}, but its use remains controversial for non-malignant pain management. Though they can be effective at managing pain in the short term, there are concerns related to its long-term efficacy, side-effects, and potential for addiction and drug abuse~\cite{rosenblum2008opioids}. Such addictive potential raises important ethical issues related to the over-prescription of opioids for chronic pain management. This is especially critical considering the current opioid abuse epidemic in the United States that was, at least in part, fuelled by drug companies encouraging and incentivizing the prescription of opioids for pain management by emphasizing benefits and downplaying the risk of addiction. It has been estimated that more than 450,000 died of opioid overdoses between 1999 and 2018 in the United States. In this context, the prescription of opioids for pain management should be done sparingly and cautiously. 

In 2020, news reported that an electronic health record (EMR) system provider that offered a cloud-based EMR solution had paid 145 million US dollars to settle a lawsuit accusing them of modifying their CDSS to promote the use of long-action opioids for the treatment of chronic pain. A review of the court documents made public by the US Department of Justice showed that the company added a treatment option, unsupported by clinical evidence, to the options presented to clinicians when deciding the next therapeutic steps for patients suffering of chronic pain. The documents revealed that the software provider solicited remuneration ``to design the Pain CDS to cause healthcare providers to extend the duration of ERO [extended-release oxycodone] prescriptions, convert patients receiving IROs [immediate-release oxycodone] to EROs, to increase the overall market of ERO-using patients, and to measure its ability to deliver such results''\cite{justice2020}.

\begin{figure}[h!]
  \includegraphics[width=80mm]{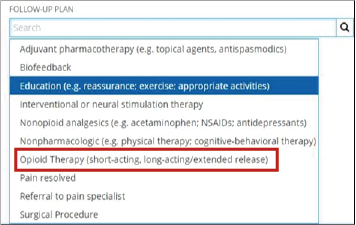}
  \caption{Drop down options presented to clinicians when designing a follow-up plan for patients with chronic pain. The option labelled \textbf{Opioid Therapy (short-acting, long-acting/extended release)} is not supported by current clinical evidence and is considered detrimental (emphasis is ours). Source: US Department of Justice}
  \label{fig:DropDown}
\end{figure}

In Figure \ref{fig:DropDown} we can see evidence presented in the legal documents showing the options presented to clinicians when defining a follow-up plan for patients with chronic pain (we highlighted the controversial option in red). The options presented are frequently used to treat patients with chronic pain, however, the company inserted an additional option suggesting the use of opioids, including long-acting an extended release opioids. The options presented were sourced from a paper published in the New England Journal of Medicine (NEJM), a highly prestigious medical journal. However, not only the paper did not suggest the use of opioids in these cases, the paper advocated \textit{against} the use of opioids given the high potential to develop opioid dependence and addictions \cite{volkow2016opioid}. From the evidence presented, it is clear that the additional option inserted was not highlighted in any way that would allow the clinician to identify its different provenance.

To further nudge clinicians to prescribe opioids, the system included three alerts. The first prompted doctors to record a pain score from the patient. The second prompted them to collect a BPI score (brief pain inventory) of patients who reported a pain score equal or higher than 4 (out of 10) twice or more in the past 3 months. The third prompted the creation of a pain management plan for the patients who reported a pain score equal or higher than 4 within four months and patients with chronic pain. The legal documents reveal that marketing professionals were involved in the design of the system, and that they believed that the questions in the BPI would focus clinicians' attention on pain symptoms and would increase the likelihood of them creating a pain management plan for the patient.

The legal documents we reviewed also present extended email chains describing the collaboration between an opioid-producing drug company and the EMR vendor. Two aspects are salient from the email discussions. First, the drug company specifically targeted ``opioid-na\"{i}ve patients'', meaning patients that had not previously used the treatment and were at high risk of abuse. Second, the EMR company produced an internal analysis that it shared with the drug company showing that extended release opioids were among the least effective interventions to reduce pain. A preliminary analysis by the EMR company explained that the CDSS had generated alerts during 21 million patient visits for 7.5 million unique patients and almost 100,000 healthcare providers after only 4 months of going live. The CDSS alert operated until the spring of 2019 after alerting more than 230,000,000 times and resulted in tens of thousands of additional opioid prescriptions.

\section{Addressing Dark Patterns in Clinical Decision Support Systems}

The case above highlights the serious consequences that dark patterns can produce in medical systems. This is an example of the \textit{interface interference pattern}---``manipulation of the user
interface that privileges certain actions over others.''~\cite{gray2018darkpatterns}---where commercial deals between the EMR provider and a pharmaceutical company led an option that is harmful to patients to be presented as an alternative that is equivalent to more suitable ones. It can also be understood as an instance of the \textit{sneaking} pattern---``attempting to hide,
disguise, or delay the divulging of information that is relevant to the user''~\cite{gray2018darkpatterns}---as the system is not upfront as to what is known about the limitations and risks of the treatment, as well as as to the commercial deals behind the recommendation. 

The effectiveness of the design decision in advancing the pharmaceutical company's goals can be explained by their misuse of several cognitive biases, including:
\begin{itemize}
    \item \textbf{Salience bias}: By designing alerts that focused the clinician's attention on the pain experienced by the patient, the system made it more salient in the doctor-patient interaction, and therefore, more likely that the doctors would work towards a pain management plan, many of which would include the prescription of opioids.
    \item \textbf{Authority bias}: The options listed in the suggestions for a follow-up plan were supposedly derived from a prestigious academic publication, which increased the likelihood that the clinician would attribute a greater accuracy to the recommendation.
    \item \textbf{Hyperbolic discounting bias}: Because people tend to prefer immediate payoffs relative to later payoffs, the short-term effectiveness of opioids makes them a more attractive option than alternatives that are more effective in the long-term.
    \item \textbf{Automation bias}: The natural tendency for users to over-rely on automation means that recommendations provided by CDSSs are often accepted with little scrutiny.
\end{itemize}

The example shows that even relatively simple interface design decisions can lead to disastrous health outcomes. This trend is likely to continue as the complexity of the models behind these decision support systems increases. As these systems begin to incorporate black box machine learning models, the tendency is for their recommendations to become even more inscrutable. Further, the use of embodied intelligent interactive agents will make these recommendations even more persuasive, requiring additional effort from the part of clinicians to overcome the cognitive biases underpinning dark patterns.

Completely addressing the problems highlighted in this case requires multiple perspectives. Here we propose a few ways in which design can contribute:

\begin{itemize}
    
    \item \textbf{Explainability}: systems should be able to explain how they came to the conclusions that they did in a way that humans can understand it. Though work has started in this area---especially in the health domain (e.g. \cite{lim2019xai})---this is a an area that requires substantially more research, both in terms of developing effective explainable algorithms and in designing human-centred explanations understandable by their end-users.
    
    \item \textbf{Knowledge Provenance}: as health information systems keep evolving, they not only contain patient information but also biomedical knowledge. Explicit methods to convey the source of external knowledge being delivered through EMRs is critical \cite{adler2019preparing}. The recent availability of APIs to interact with EMRs \cite{watkins2020fhir} should open the possibility to integrate computable knowledge from trusted and verifiable external sources. 
    

    \item \textbf{Educating health professionals} about the limitations of automated decision making systems: people often believe in recommendations from AI systems as if they were true (e.g. \cite{wouters2019biometric}). There should be more awareness about the limitations of these systems, both in terms of their general limitations (e.g. data bias, overfitting, etc.) and within the applications themselves (e.g. visualising the uncertainty of specific recommendations, tracing the provenance of the recommendations in a visual manner).

    \item \textbf{Libertarian paternalism:} any interface design will privilege some alternatives over others. The example above shows that even though the system did not overtly highlight the opioid option, the simple presence of that alternative was problematic. Designers must acknowledge cases where the options are not equivalent and create choice architectures that privilege consensus opinions and well-chosen defaults. The libertarian paternalistic approach is one that preserves the agency of the clinician but that nevertheless nudges them towards  directions that will promote the welfare of the patients~\cite{thaler2003libertarian}. 
\end{itemize}

Addressing dark patterns in the design of CDSSs is a complex and wicked design problem, requiring expertise from health professionals, user experience and interface designers, cognitive psychologists, digital ethicists, among others. Ultimately, it is critical that the system supports clinicians to make decisions that explicitly prioritise patients' interests and outcomes.  



\bibliographystyle{unsrt}
\bibliography{bibliography}


\end{document}